# Magnetohydrodynamics and magneto-solutal transport mediated evaporation dynamics in paramagnetic pendent droplets under field stimulus

**Vivek Jaiswal [1], Raghvendra Kumar Dwivedi [1], A R Harikrishnan[2] and Purbarun Dhar [1, *]**

[1] Department of Mechanical Engineering, Indian Institute of Technology Ropar, Rupnagar–140001, India

[2] Department of Mechanical Engineering, Indian Institute of Technology Madras, Chennai–600036, India

*Corresponding author*: E–mail: purbarun@iitrpr.ac.in ;pdhar1990@gmail.com

Phone: +91–1881–24–2173

## Abstract

Evaporation kinetics of pendant droplets is an area of immense importance in several applications in addition to possessing rich fluid and thermal transport physics. The present article experimentally and analytically sheds insight into the augmented evaporation dynamics of paramagnetic pendent droplets in the presence of a magnetic field stimulus. Literature provides information that solutal advection and solutal Marangoni effect lead to enhancement of evaporation in droplets with ionic inclusions. The major crux of the present article remains to modulate the thermo-solutal advection with the aid of magnetic field and comprehend the dynamics of the evaporation process under such complex multiphysics interactions. Experimental observations reveal that the evaporation rate enhances as a direct function of the magnetic moment of the solvated magnetic element ions, thereby pinpointing at the magnetophoretic and magneto-solutal advection. Additionally, flow visualization by Particle Image Velocimetry (PIV) illustrates that the internal advection currents within the droplet are strengthened in magnitude as well as distorted in orientation by the magnetic field. A mathematical formalism based on magneto-thermal and magneto-solutal advection effects has been proposed via scaling analysis of the species and energy conservation equations. The formalism takes into account all major governing factors such as the magneto-thermal and magneto-solutal Marangoni numbers, magneto-Prandtl and magneto-Schmidt numbers and the Hartmann number. The modeling establishes the magneto-solutal advection component to be the domineering factor in augmented evaporation dynamics. Accurate validation of the experimental internal circulation velocity is obtained from the proposed model. The present study reveals rich insight on the magneto-thermo-solutal hydrodynamics aspects in paramagnetic droplets.

# 1. Introduction

Fluid dynamics, thermal and species transport within and from microliter droplets is emerging as an established field of research since the past few decades owing to the critical role microdroplets play in different systems, applications, and processes. Broadly speaking, important applications of droplets and sprays(a populous collection of microdroplets)[1-3] include the automobile sector, propulsion systems, combustors [4-6], condensers, cooling towers and HVAC systems, etc. and thus clarity regarding the fluid dynamics, heat transfer and evaporation/ condensation of micro-scale droplets is essential to optimize the performance of such systems. Other applications where droplets play vital roles are biomedical systems, like patterning and detection of diabetes and blood and pathological tests, drug delivery through inhalers, nebulizers and spray based painkillers or fumigants [7], etc. Improvements of pesticides or insect repellents spray [8], ink-jet printing technologies, spray cooling and humidification, processing and manufacturing [9] and domestic HVAC are adding importance to research in the field of fluid mechanics, heat and mass transfer from droplets.

Standalone droplets are in general categorized into two types, the pendant, and the sessile droplets. While the latter rests in equilibrium on a solid surface and surrounded by the ambient gas phase; the former suspends in equilibrium from the tip of the solid surface (often a needle or thin structures) with the ambient gas phase shrouding the majority of the droplet surface. In addition to the equilibrium of phases, surface properties like wettability and surface energy also play vital roles in case of the sessile droplets. Some well documented studies involving thermofluidics of sessile droplets are reports by Hu and Larsen [9], Semenov et al. [10], Bekki et al. [11] and Fukai et al. [12], etc., where the crux of the studies were concentrated upon the various interfacial dynamics associated with the evaporation kinetics of such droplets. In the area of droplet thermofluidics, the major focus has been on exploring the dynamics of sessile droplets, but recent trends in research show a paradigm shift towards pendent droplets, which is mostly influenced by the fact that pendent droplets are independent of the surface interaction that sessile droplets experience. Pioneering investigations on the evaporation dynamics of fuel based pendant droplets was initially reported by Godsave [13], and molecular diffusion to the ambient was argued to be the driving parameter for the observed evaporation process. Based on the analysis of the ambient phase, the classical $D^2$ law was presented as a simple mathematical representation of the evaporation process. The law assumes spherical droplet, with constant diffusion rate, evaporating in a quasi-steady gas phase in an isobaric manner, and with uniform droplet temperature. Over the years, seminal studies on pendent droplets have evolved, such as by Kuz [14], where reports confirm compliance with the $D^2$ law and with Hertz- Knudsen model, which relates thepressure with the time rate of change of concentration of molecules during evaporation.

Fluid dynamics, thermal and species transport phenomena in conducting or magnetically active fluids in presence of magnetic field (commonly termed as magnetohydrodynamics), has been an area of sincere interest within the academic community given the rich physics of the problem. In general, such flows are characteristics of geophysical flows (such as mantle convection, flows within the magnetosphere, etc.), or in astrophysical flows (such as the convective zone within stars, gas flows in accretion disks, etc.). At small or man-made scales, such dynamics can be observed in systems such as plasma reactors, particle accelerators, etc. In recent times, magnetohydrodynamics has been of interest in ferrofluid based targeted drug delivery. Ferrofluids or magnetic fluids are stable solutions of magnetically active dispersed phase (mostly nanoscale iron particles in case of ferrofluids) in a carrier solution (Laroze et al. [15]), such as water, polymers, oils, and stabilized using a surfactant. The magnetically active dispersed phase could be either paramagnetic salts or magnetic nanoparticles and the behavior of the system is strongly dependent on the dispersed phase characteristics and the thermo-physical response of the system is field dependent. This behavior is caused by the orientation of the magnetic particles or the solvated paramagnetic ions along the direction of the magnetic field. In comparison to ordinary fluids, magnetic fluids exhibit two additional supplementary features; the Kelvin force and the body couple (Rosenweig [16]) in the presence of a magnetic field. The Kelvin force (or polarization force) is the force by virtue of which any magnet attracts magnetic domains within a body while the body couple is the angular momentum generated within a magnetic component due to the torque arm from the magnetic field source.

Some studies on fluid flow and heat transfer with magnetic field interaction have been reported in the literature; however, given the complexity of the problem, the vast majority of such studies are computational by nature. Rossow [17] simulated and reported several aspects of change in flow characteristics and heat transfer due to magnetic field involvement with incompressible electrically conducting fluids. For a boundary layer over an infinite flat plate with transverse magnetic field applied, determining the temperature and velocity profile was considered in the study. The magnetohydrodynamic interaction was reported to give rise to a new parameter, in the form of the ratio of electromagnetic force density to the dynamic pressure (m=$\sigma B^2/\rho U_\infty$). The product of 'm' and separation of the magnetic field and surface of fluid flow was proposed as the guiding parameter for the effectiveness of magnetic field modulated change in the rate of heat transfer. Sparrow and Cess [18] investigated free convection fluid dynamics and heat transfer along a vertical plate in presence of magnetic field. Analytical expressions show drastically modified fluid and heat flow behaviour due to the magnetic body force. Rudraiah et al. [19] also computationally studied free convection of electrically conducting fluids confined in rectangular enclosures. The effect of magnetic field was reported to bring about a reduction in the Nusselt number due to grossly varying flow dynamics at different Hartmann numbers. Alpher [20] reported change in fluid flow and heat transfer in parallel shear flows due to the presence of an external magnetic field. Magyari et al. [21-22] and Kandasamy et al. [23]compiled comprehensive details on the analytical and numerical outlooks in magnetohydrodynamics, andconcluded that fluid dynamics and thermal transport in electrically conducting or magnetically active fluids can be effectively modulated by the use of magnetic fields.

Computational analysis and stability theory has been also extended to thermal and mass transport systems in fluid flow under influence of magnetic fields. Mudhaf and Chamkha [24] and Lin et al. [25] investigated magnetohydrodynamic (MHD) adsorption and thermal gradients in order to study thermo-capillary Marangoni convection under field effect. Studies on the convective transport in ferrofluids due to density gradients have also been reported (Sunil and Mahjan [26]),where the focus was on understanding the Rayleigh-Benard convection in such fluids under magnetic field effect. Singh and Bajaj [27] and Belyaev and Smorodin [28] explored ferro-convection behavior under the stimuli of sinusoidal temperature profile and alternating magnetic field and simulation results were presented. A numerical approach was explored for the case of magnetic field dependent viscosity and its implications on Bernard ferroconvection (Nanjundappa et al. [29]) and comparison of the numerical results with analytical results (Lebon and Cloot [30]) in absence of magnetic field was used to report an appropriate numerical scheme for such non-linear problems. Thereafter, Rudraiah et al. [31] computationally discussed the effect of temperature gradients on Marangoni ferroconvection and comparison between analytical and numerical results (Nield [32]) show that the magnetic field strengthis an important parameter that determines the onset of Marangoni convection in ferrofluids.

Thereby, detailed study of the literature reveals several interesting facts. Firstly, the influence of magnetic field on fluid flows and heat transfer have been studied, but in general computationally. Secondly, the majority of such studies deal with ferro-convection, which is typical of fluid systems of high electrical conductivity in the presence of very strong magnetic fields. Thereby, the field at hand has paid very scarce attention to realistic thermofluid problems, such as the simplistic case of a paramagnetic fluid undergoing evaporation in the presence of a magnetic field. The present paper aims to understand the complex behavior and interplay of fluid dynamics and heat transport during theevaporation process of a paramagnetic droplet (achieved employing solvated magnetic salts) in the presence of anapplied magnetic field. Study of evaporation of ionic solution droplets (Jaiswal et al. [33]) has established that such droplets evaporate faster than water droplets due to augmented internal circulation caused by solutal advection. A deriving clue from the findings, it can be argued that the circulation of paramagnetic ions within the fluid in presence of magnetic field would constitute a probsimilar to the motion of a charge within a magnetic field. This would induce redistribution of the ionic concentration in the fluid, thereby changing the evaporation kinetics, which is a gist of the focus of the present article. Experimentally understanding the physics of coupled thermofluidics and magnetism in evaporating droplets could have far-reaching implications in microfluidic multiphysics.

## 2. Experimental methodologies

In the present experiments, three paramagnetic salts, viz. anhydrous iron (III) chloride ($FeCl_3$, Sigma Aldrich, India), cobalt chloride hexahydrate ($CoCl_2.6H_2O$, Merck, India) and nickel sulphate hexahydrate ($NiSO_4.6H_2O$, Merck, India), all of the analytical reagent grade, have been used. The salts were dissolved in deionized water (Millipore) and three different concentration solutions of 0.05 M, 0.1 M, and 0.2 M were used for the experiments. The relationship between concentration in molarity (M) and in kg/m$^3$ (kg of salt per m$^3$ of solvent) with respect to the molecular weight ($M_{wt}$) of the salt is expressed in eqn.1.

$$Concentration(kg/m^3) = M_{wt}.Concentration(M) \qquad (1)$$

The experiments were performed using a customized setup and figure 1 illustrates the schematic of the experimental setup. A droplet dispensing mechanism with a precision digitized controller (Holmarc Opto-mechatronics Ltd., India) is employed to generate the pendent droplet of required volume. A sterile glass syringe dispenses the fluid which hangs from the tip of the needle (stainless steel, flat ended) by virtue of its surface tension. The syringe has a least count of 0.1 µL and the volume of the droplets used in the present experiments is 25 ± 0.5 µL throughout. The diameter of the pendant droplet is observed to be 2.8 ± 0.2 mm, in general. The magnetic field was generated across the droplet by employing a horizontal pole electromagnet setup with digitized current based control (Polytronic Corporation, India). The droplet was suspended vertically from the needle exactly at the center of the pole pieces of the electromagnet such that the field lines pass horizontally through the droplet. The distance between the poles has been maintained at 20 mm such that the poles are far enough from the droplet to otherwise influence evaporation kinetics. The electromagnet was initially calibrated (at pole separation of 20 mm) against current input and constant voltage using a digitized Gaussmeter (with an InAs based sensor). At the pole spacing of 20 mm, the electromagnet is capable of achieving 6000 Ga field strength. For the present experiments, three magnetic field intensities, viz. 800, 1600 and 2400 Ga have been used. It has been observed that for 3000 G and beyond, the magnetic poles exhibit mild Joule heating. This is potent enough to thermally influence the evaporation dynamics drastically and hence, the field strength has been limited to 2400 Ga (0.24 T).

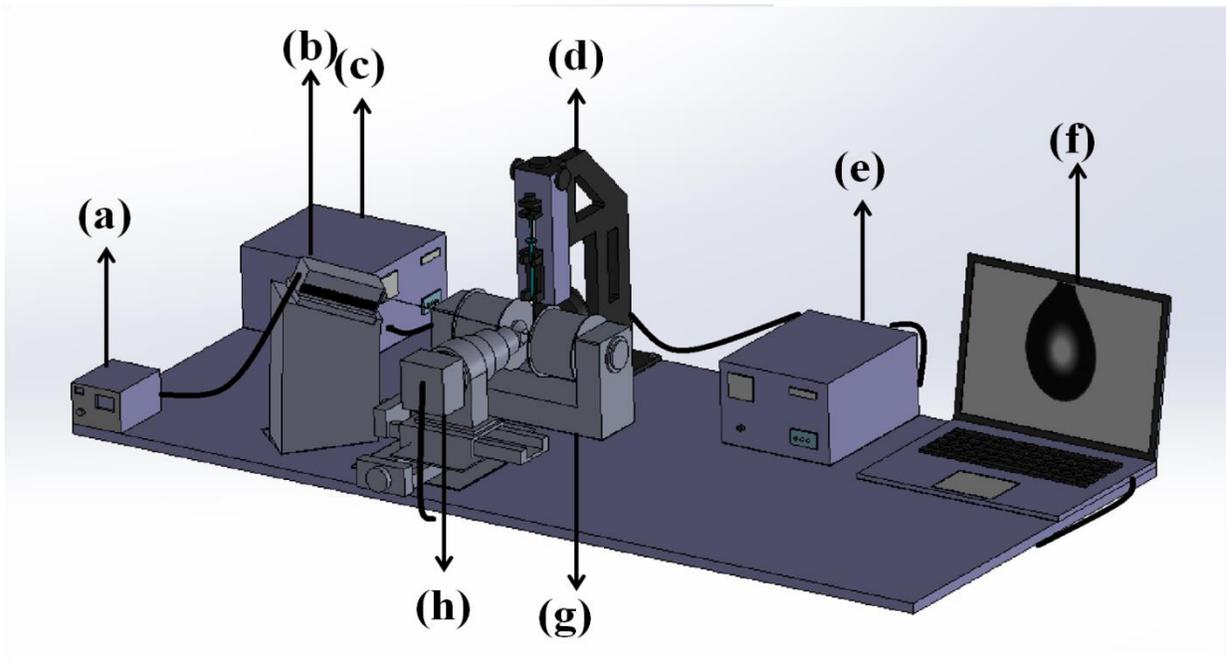

**FIG. 1** Schematic of the experimental setup employed (a) Laser controller (b) Laser mounted on stand (c) Electromagnet power supply and controller (d) Droplet dispensing mechanism (DDM) with backlight illumination attachment (e) DDM and illumination controller (f) Data acquisition and camera control computer (g) Electromagnet unit (h) CCD camera with microscopic lens assembly and two-axis movement control. Components (b), (d), (g) and (h) are enclosed in an acrylic chamber.

The evaporation process was recorded using a CCD camera (Holmarc Opto-mechatronics Ltd., India) with a microscopic lens assembly attached to it. The resolution of the camera is 1280 x 960 pixels at a recording rate of 10 fps (used presently). A brightness controlled LED array light source was positioned opposite to the camera behind the droplet as an illumination source. The complete setup was housed within an acrylic chamber and placed on a vibration free horizontal board to arrest all potential ambient disturbances. The measurements of ambient temperature and humidity were performed using a digital thermometer and a digital hygrometer respectively, and both were placed inside the chamber. Both devices contain probes to measure the local temperature and humidity, and these probes were positioned 20 mm away from the droplet. During the experimental runs, the temperature and humidity were observed as 25 ± 1$^o$C and 50 ± 4 % throughout. The recorded images of the evaporation process have been processed using the open-source software ImageJ. A macro was written for batch processing the array of images to obtain required geometric parameters. The images were first converted into the binary color scheme and then the droplet profile was fit to the closest ellipsoid. This was further fit to the equivalent sphere to obtain the equivalent diameter for the instantaneous pendant profile. The evaporation rate of deionized water serves as the benchmark and at zero magnetic fields, the evaporation rate value confirms to reported literature (Mandal and Bakshi [34], Jaiswal et al [33]), thereby validating the customized experimental setup. Internal flow dynamics within a droplet often play significant roles in modulating kinetics of evaporation [33]. Visualization of flow features and quantification of velocities were realized using Particle Image Velocimetry (PIV). Neutrally buoyant (with water at 300 K) fluorescent particles (polystyrene, 10 μm diameter) were used as seeding particles (Cospheric LLC, USA). A continuous wave laser (Roithner GmbH, Germany) of wavelength 532nm and 10 mW peak power is used as the illumination source for the PIV studies. A laser sheet of ~1 mm thickness using a plano-convex lens has been employed for illuminating the droplet interior. The PIV images are recorded for 1 min during the first few minutes of the evaporation process (the whole droplet takes up nearly 40–45 minutes to evaporate). During the PIV image acquisition, the backlight was dimmed and the laser was employed as the illumination source. The camera resolution used for the PIV was ~ 120pixels/mm and recording was done at 30 fps. For post-processing, a cross-correlation algorithm with multiple passes of 64 pixels, 32 pixels, and16pixels interrogation windows was used to harness high signal to noise ratio in the velocity determination process. The data was processed for typically 300 images for each case employing the open source code PIV lab. The average displacement of the particles is a function of the salt concentration and strength of magnetic field applied, however, for a typical 0.1 M $FeCl_3$ based droplet, the average displacement of particles between frames for zero fields was observed to be ~ 0.3 mm.

## 3.Results and discussions

### 3.1. Evaporation characteristics in presence of a magnetic field

The following sections discuss the physics and mechanisms of evaporative dynamics of paramagnetic pendent droplets in conjunction with magnetic fields. Figure 2 (a) and 2 (b) illustrates the evaporation kinetics of droplets of water and all three salt solutions (0.1 M) in the absence of afield and at 1600 Ga (0.16T) field strength. It is observed that the classical $D^2$ law (Eqn. 2) holds valid even in the presence of magnetic field, however, the slope of the lines are observed to increase in the presence of magnetic field.

$$\frac{D^2}{D_0^2} = 1 - k\frac{t}{D_0^2} = 1 - k\tau \tag{2}$$

In eqn. 2, D is the instantaneous equivalent spherical droplet diameter, $D_0$ is the initial diameter of the droplet, k is evaporation rate and t is the elapsedtime duration of evaporation. Figure 2 (a) serves as the datum or reference for zero magnetic field case. Comparing Figure 2 (b) with (a), the effect of magnetic field on the evaporating dynamics of the droplet is brought forward. Figure 2 (a) and (b) both illustrate that iron chloride based droplet exhibits the highest evaporation rate among all three salts, followed by cobalt chloride hexahydrate and nickel sulphate hexahydrate. Further evident from Figure 2 (a) is the fact that even in the absence of a field, the presence of salt enhances the rate of evaporation. The increment in evaporation of droplets due to dissolved salts has been reported by Jaiswal et al. [33]. It has been shown that the presence of salts induces thermo-solutal advection within the droplet (due to concentration gradient within the droplet) and the interfacial shear replenishes the diffusion layer of vapor shrouding the droplet, thereby enhancing mass transport from the droplet to the ambient. It may be further observed from Figure 2 (a) that even at the same concentration; different salts are able to enhance evaporation rates by different amounts. This has also been explained [33] and theoretically shown that the increment in evaporation rate is a direct function of the solubility of the salt. As evident from Table I, the evaporation rate holds a direct relationship with the solubility of the salt.

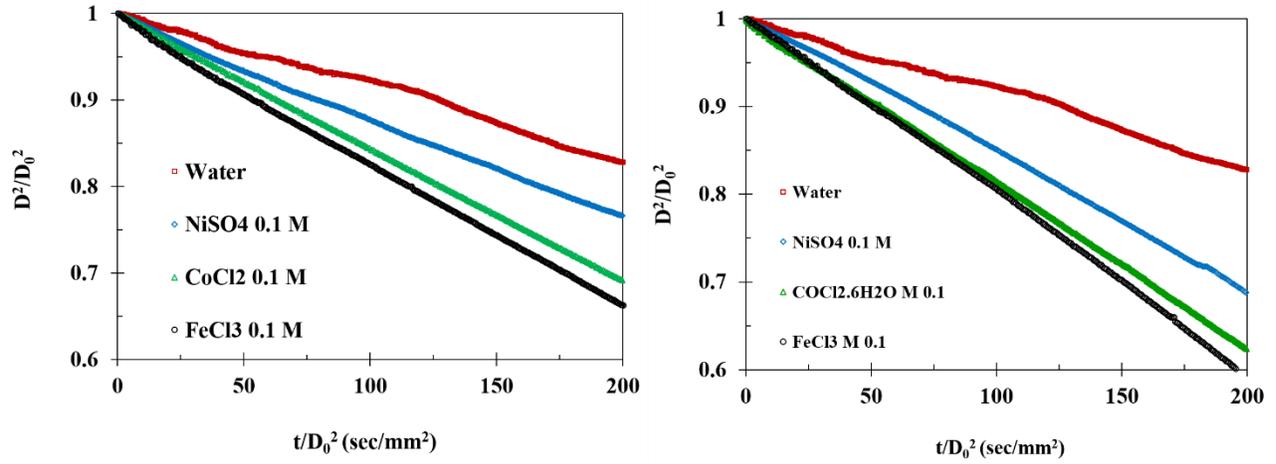

**FIG. 2** Evaporation characteristics of water and different magnetic salts (FeCl$_3$, CoCl$_2$.6H$_2$O, and NiSO$_4$.6H$_2$O) based droplets of 0.1 M concentration at (a) 0 Ga (b) 1600 Ga 0.16 T.

**Table I:** Solubility, magnetic moment of the cation (at 30-40°C) and evaporation rates for different salt based droplets (at zero and 0.24 T fields denoted as k$_{0T}$ and k$_{0.24T}$ respectively) at 1 atm pressure

| Salt | Solublity (gm/100 mL) | Evaporation rate constant k$_{0T}$ (mm$^2$/sec) | Magnetic moment (N-m/T) | Evaporation rate constant k$_{0.24T}$ (mm$^2$/sec) |
|---|---|---|---|---|
| FeCl$_3$ | 107 | 9.45 x 10$^{-4}$ | 0.8313 | 1.87 x 10$^{-3}$ |
| CoCl$_2$ | 59.7 | 9.41 x 10$^{-4}$ | 0.3265 | 1.73 x 10$^{-3}$ |
| NiSO$_4$ | 46.6 | 9.32 x 10$^{-4}$ | 0.1589 | 1.65 x 10$^{-3}$ |

It is observed from Figure 2 (b) that the evaporation rates for the salt based droplets enhance in the presence of magnetic field; however, the water case remains largely intact. Guo et al. [35] discussed the enhancement of evaporation rate of water droplets in a magnetic field due to the difference in the magnetic susceptibility between water vapor and surrounding air. At very strong magnetic fields (of the order of 10 T), the intrinsic diamagnetism of water leads to the generation of repulsive forces within the droplet, which aids in weakening the hydrogen bond of water molecules, thereby facilitating the departure of water molecules from the surface [33]. However, in the present scenario the field strengths are one order of magnitude smaller, and thus no effect of magnetic field is observed in the water droplet case. Figure 2 (b) presents another interesting aspect that different salts exhibit different evaporation rate under the influence of magnetic field. It has been further observed that there is a possible additional factor in addition to the solubility of salts which differentiates evaporation of different salt solutions under magnetic field effect. The analysis shows that just like the degree of change in evaporation rate at zero fields has a direct relationship to the solubility, the same under field effect holds a strong correlation also to the magnetic moment of the associated magnetic cation. Table 1 also exhibits that larger value of the magnetic moment of the paramagnetic cation induces larger enhancement in evaporation rate under field influence. Thereby, it can be argued that the presence of these typical cations (themselves paramagnetic with high positive magnetic susceptibility) in the droplet induces paramagnetism within the fluid, which could possibly play a role in the enhanced evaporation kinetics and will be discussed later in depth. Figures 3 (a) and (b) presents a time snap array of the evaporating droplet with different salt concentrations and under varying magnetic field intensity. The influence of different magnetic salts and magnetic field strengths produces observable change in the life time of droplet. Interestingly, the presence of hydration molecules in certain salts may aid the evaporation process, however, that is restricted to the very initial regime (Charlesworth and Marshall [36]) and therefore not of interest in the present scenario.

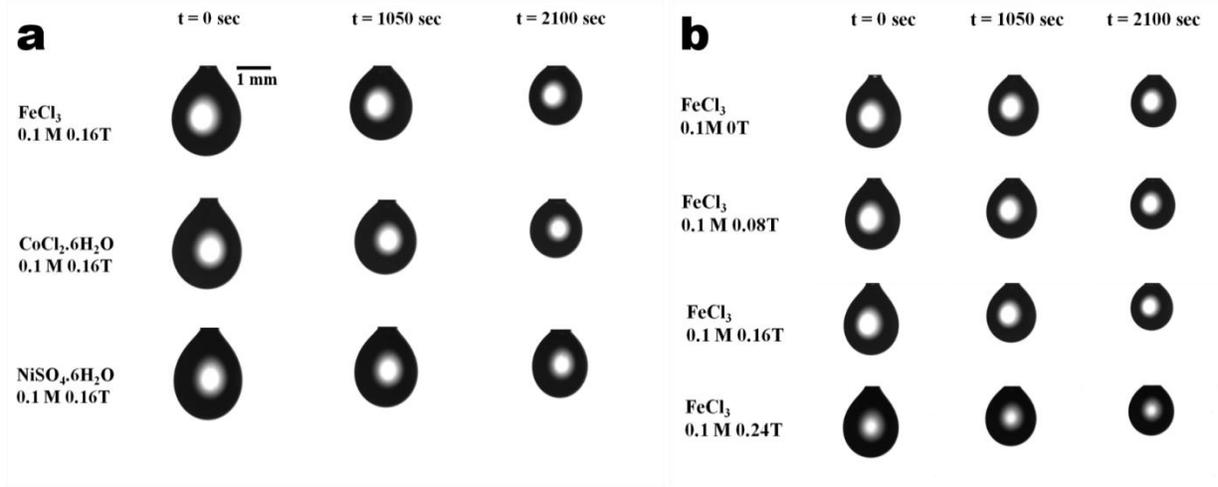

**FIG. 3** Time snaps of (a) different salts at 0.1 M concentration and 0.16 T field and (b) different magnetic field strength for $FeCl_3$ at 0.1 M. The scale bar corresponds to 1 mm.

Figure 4 (a) illustrates the changes brought about in the evaporation process by variant magnetic field strengths for a droplet of a $FeCl_3$ solution of 0.2 M concentration. Figure 4 (b) illustrates the evaporation rate constants at different magnetic field strengths, obtained from the $D^2$ law applied to the data in Figure 4 (a). The 0 T case in Figure 4 (a) provides the datum for evaluation of change in evaporation dynamics in presence of magnetic field. Since the salt and initial concentration is same, the solutal and solubility effects [33] are similar, and hence changes observed are solely brought about by the magnetic field strength. The 0.08 T case does not exhibit any typical difference from the 0 T and this is possibly due to the low paramagnetic moment imparted to the fluid system at low field strengths. At 0.16 T, the effective paramagnetic moment of the fluid phase enhances and could be responsible for the improvement of the evaporation rate by 15%. The 0.24 T case shows major improvement in evaporation rate by 50% from the 0 T case and the enhanced magnetic moment could be a responsible factor. However, as discussed earlier, the susceptibility mechanism is not valid for the present case, as the field strengths are one order of magnitude smaller than that required to induce susceptibility gradient driven enhanced molecular diffusion from the droplet. Accordingly, plausible physics for the enhanced magneto-evaporation dynamics needs to be established.

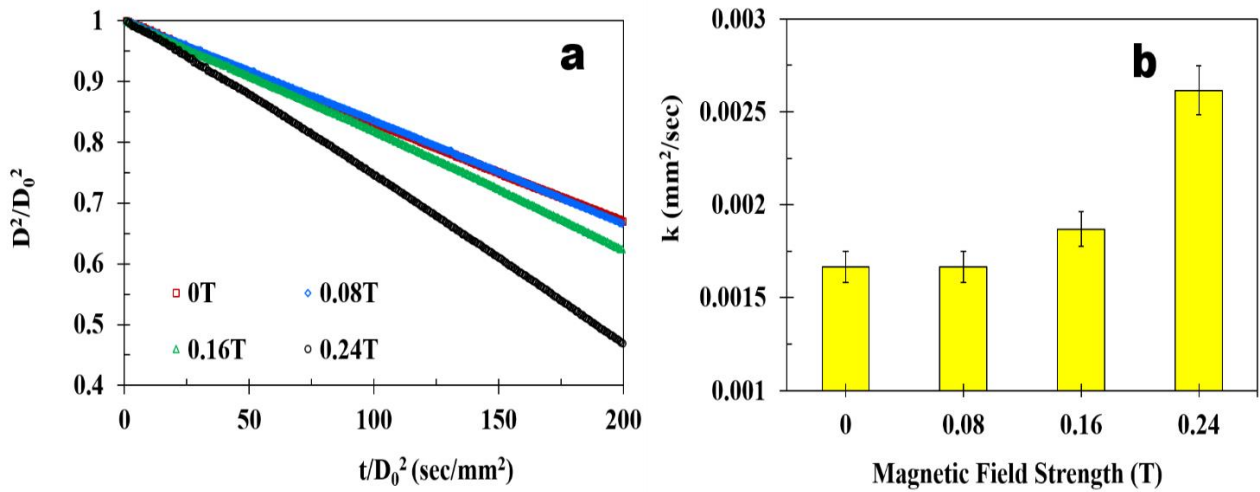

**FIG. 4** (a) Evaporation characteristics of a $FeCl_3$ solution of 0.2 M under the influence of different magnetic field strengths (b) Evaporation rate constants corresponding to magnetic field variation.

## 3.2. Role of surface tension change under magnetic field influence

Surface tension plays an important role in the thermofluid dynamics of pendent droplets as the pendant hangs stably against gravity by virtue of surface forces at the needle– fluid interface. When the magnetic field is applied against such magnetically active fluids, it is expected that the paramagnetic ions in these typical salts will tend to align or orient to the direction of the magnetic field, and this may affect the physical as well as chemical transport properties of the solution. Typically, studies have shown that hydrated ions in salt solutions tend to desorb away from the surface towards the fluid bulk, thereby leading to change of the effective surface tension compared to water [33]. Figure 5 illustrates the variation of surface tension under the influence of different magnetic field strengths for different concentration solutions of cobalt chloride. The surface tension value of the deionized water sample employed is ~ 69 ± 2 mN/m. It is observed that while the minor change in the value of surface tension is brought about by the concentration of the salt, there is also no significant change due to increasing magnetic field strength. The variation of surface tension due to field effect ranges within the uncertainty range of ± 2 %, which is not potent enough to induce drastic increment or decrement in evaporation. Figure 5 (b) illustrates the evaporation rate constants for different salt solutions at 0.2 M concentration under the influence of magnetic field variation. As discussed in the earlier section, the salt solutions exhibit enhanced evaporation dynamics in tune with their effective paramagnetic moment imparted by the magnetic cations. Thereby, the nickel sulphate based salt exhibits the weakest improvement in evaporation owing to its low value of the magnetic moment. However, the increment in evaporation rate at 0.24 T is ~ 3 times compared to the zero field case. This large enhancement cannot be explained by the minimal changes in surface tension under field effect and hence additional experimental support is required to understand mechanism completely.

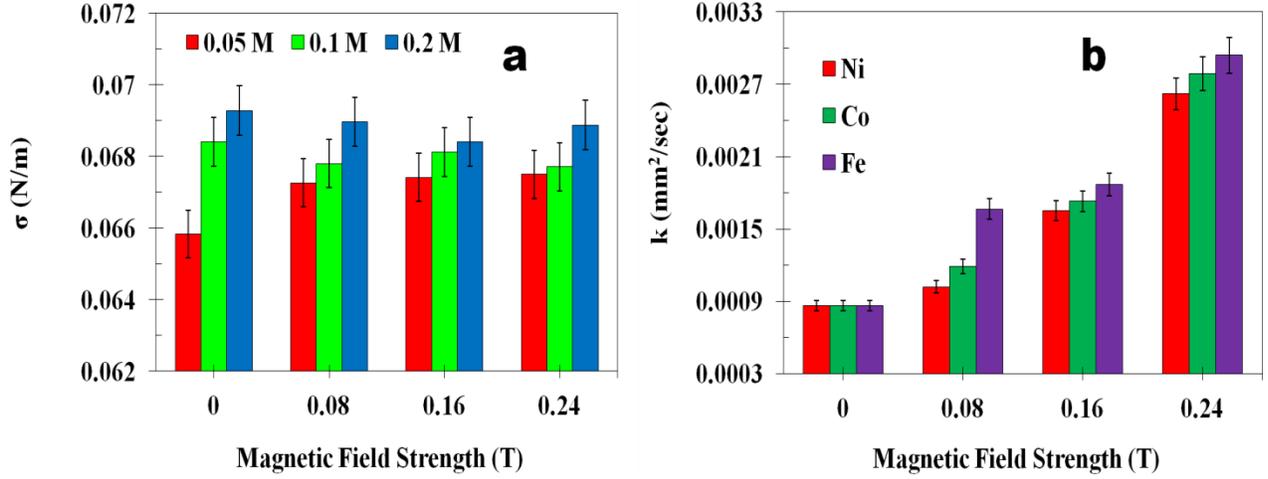

**FIG. 5** (a) Surface tension variation for CoCl$_2$.6H$_2$O based solution as a function of magnetic field strength for different concentrations (b) behavior of the evaporation rate constant for different salts at 0.2 M concentration as a function of magnetic field strength.

### 3.3. Diffusion dominated evaporation dynamics under magnetic field effect

Given the fact that role of surface tension change has been removed from consideration as a potential mechanism to explain the enhanced evaporation, further possible mechanisms must be worked upon. Evaporation in case of pure fluids in absence of ambient advection is by molecular diffusion [13]. The diffusion driven evaporation kinetics is mathematically represented by the classical model of Abramzon and Sirignano [37] and allows determination of the evaporation rate. The model [37] determines the driving gradient due to vapor concentration difference across the diffusion layer around the droplet and the ambient air and deduces the evaporation rate as per Eqns. 3–5

$$B_m = \frac{Y_s - Y_\infty}{1 - Y_s} \tag{3}$$

$$\frac{dm_d}{dt} = -\dot{m} \tag{4}$$

$$\frac{dD^2}{dt} = \frac{4\rho_g D_v}{\rho_l} \ln(1 + B_M) = k \qquad (5)$$

In eqns. 3-5, $B_m$ represents Spalding mass coefficient or mass number, $Y_\infty$ the mass fraction of vapour in ambient, $Y_s$ the mass fraction of vapour in the diffusion layer shrouding the droplet, $D_v$ the diffusion coefficient of vapour with respect to the ambient gas phase, $\rho_g$ and $\rho_l$ the density of ambient gas and density of liquid and k is the evaporation rate. The terms $D$, $\dot{m}$ and $m_d$ represent the instantaneous diameter of the droplet, mass evaporated and instantaneous mass of droplet. Figure 6 (a) illustrates the theoretically deduced evaporation rate constant from the classical model [37] in comparison with the experimental values for different salts and different field strengths. It is observed that the diffusion driven evaporation model falls short to explain the enhanced evaporation kinetics. This essentially suggests that the diffusion process remains unhampered and additional mechanisms are at play for the observed increment. It is noteworthy that even at zero fields; the diffusion model falls short of explaining the evaporation of salt solution droplets. However, the evaporation dynamics of salt solution droplets has been explained in the literature [33] and the present focus shall be to explain the further enhancement under field stimulus.

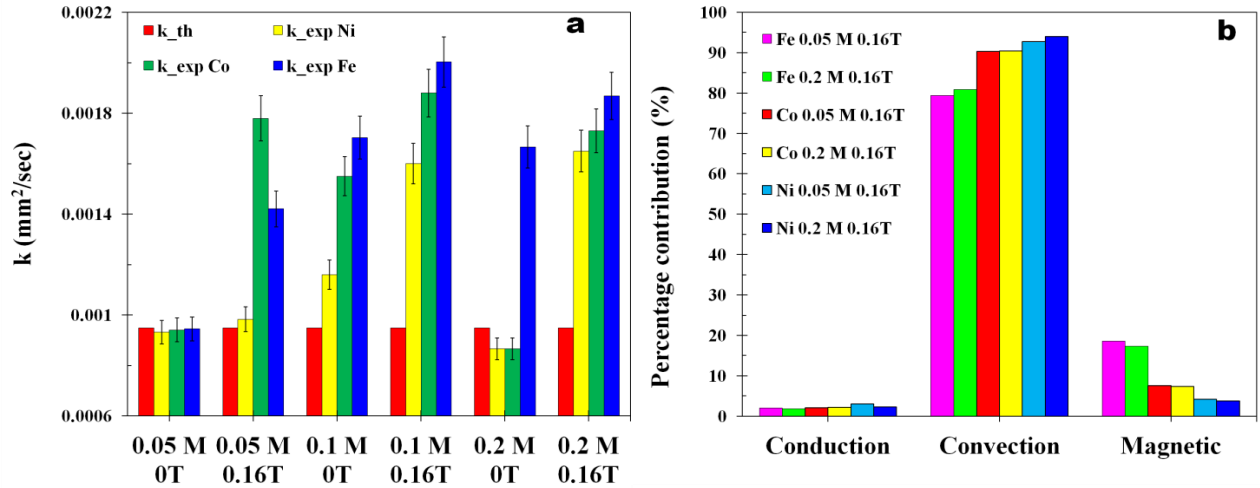

**FIG. 6** (a) Comparison of theoretical and experimental evaporation rate constant for different salts with varying concentration and magnetic field employing the diffusion model (b) Percentage contribution of the different mechanism involved in energy transfer based on the present analysis.

### 3. 4. Role of enhanced internal advection

It has been reported that solutes or inclusions change evaporation dynamics of droplets [38] and presence of augmented internal circulation within the droplet [34, 35] is a major responsible factor. PIV studies were performed to ascertain and quantify the internal circulation dynamics and were performed during the very initial stages of evaporation to avoid effects of changes in the bulk concentration of the solute phase within the evaporating droplet. The PIV was performed during the first 5 minutes of evaporation of respective experiment wherein the bulk concentration of the solute phase has been determined toincrease within the limit of 10% of the original value. The PIV image acquisition was performed at 10 fps for a period of 60 seconds and contour plots were obtained based on the averaged spatial velocity distribution for the whole data set. Figure 7 (a) and (b) illustrate the velocity contours and vector fields for 0.1 M $FeCl_3$ solution at 0 T and 0.16 T, and figures 7 (c) and (d) illustrate 0.2 M FeCl$_3$ solution at 0.08 T and 0.24 T, thereby covering concentration well as field effects. For the case of evaporation of water, no persistent internal circulation was observed; however, minor drift with velocities < 0.1 cm/s was noted. This observation validates the current PIV setup since reports [33, 34] discuss such minimal drift within water droplets with velocities ≤ 0.1 cm/s. Figure 7 (a) illustrate the contour and velocity vector plot for 0.1 M FeCl$_3$ in absence of magnetic field, and circulation pattern is observed to exist within the droplet. The presence of ionic inclusions in the droplet has been reported to give rise to solutal advection within the fluid; thereby leading to circulation [33], and the present case is in compliance to the report.

During the PIV, the illuminating laser is positioned so as to coincide with the plane of dominant circulation to the best possible extent, such that accuracy in determining the average circulation velocity is sufficient. In the presence of magnetic field, it is typically observed (from figure 7 (b)) that the plane of dominant circulation shifts and the circulation becomes dominant about the vertical axis perpendicular to the magnetic field lines. Accordingly, the average circulation velocity and contours are determined via repositioning of the laser and imaging camera. Additionally, the overall magnitude of the circulation velocity is observed to increase as a function of field strength. The reorientation of the circulation plane can be explained based on the principle of Fleming's rule. In the presence of a magnetic field, the flow of charge (due to the ionic circulation within the droplet) experiences a force orthogonal to both its own direction and the magnetic field. Thereby, the ionic circulation is reoriented in the new plane. Furthermore, the paramagnetic ions within the fluid respond to the magnetic field orientation and experience additional drift velocities, thereby augmenting the overall circulation velocity as well. It is evident from the visualization studies (Figure 7) that the effective magnitude of the circulation velocity enhances with magnetic field strength. This internal circulation has been established to be caused by thermo-solutal advection [33, 34] in case of solute based droplets or multi-component droplets, and the major discussion is present in these reports in the literature [34,35]. In brief, the thermo-solutal gradients generated within the droplet leads to advection currents, which manifests as internal circulation. This internal circulation induces shear on the droplet interface, which further imparts shear onto the vapor diffusion layer shrouding the droplet surface. This shear driven flow within the vapor diffusion layer replenishes the layer with ambient air, thereby replenishing the vapor concentration gradient across the diffusion layer. This renewed gradient ensures enhanced droplet evaporation rate. Thereby, the observation of enhanced circulation velocity under field effect points towards the fact that there is a possibility that the field modulates the thermo-solutal gradients in a manner so as to enhance the solute concentration gradient or the thermal gradients within the droplet. This would lead to enhanced thermal or solutal advection and would explain the enhanced velocity, and thus needs to be discussed further.

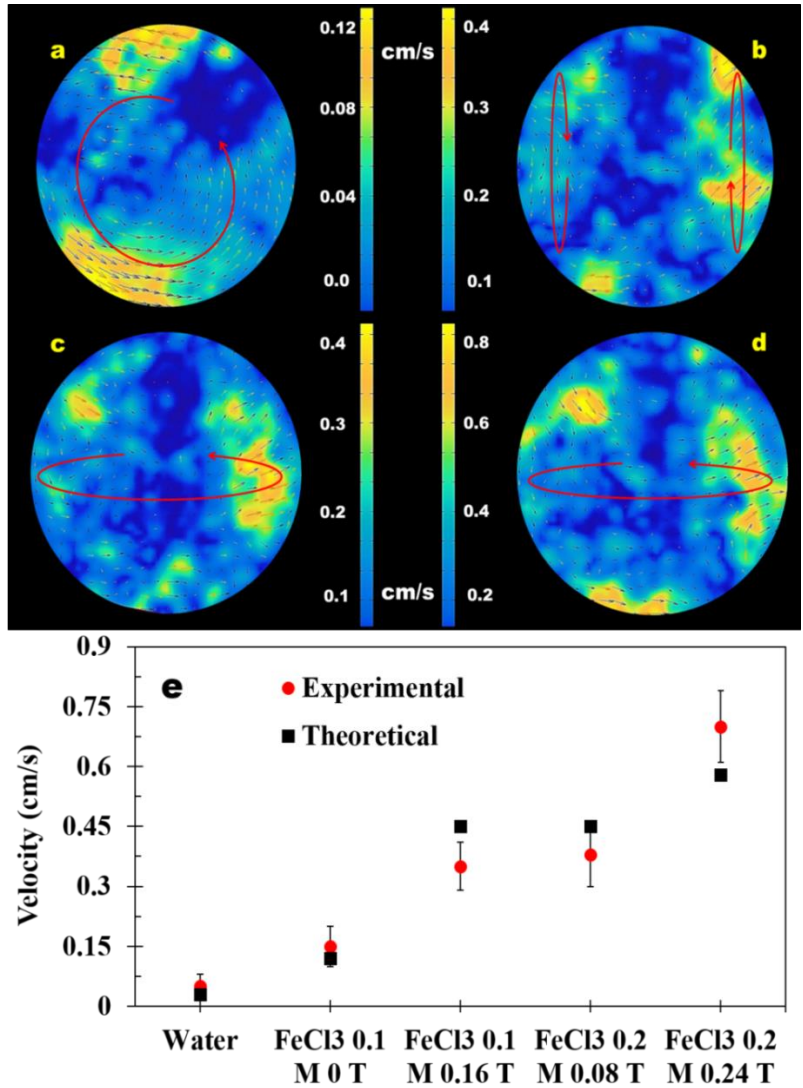

**FIG. 7** Velocity contours and vector fields for (a) 0.1 M FeCl₃ at 0 T (b) 0.1 M FeCl₃ at 0.16 T (c) 0.2 M FeCl₃ at 0.08 T (d) 0.2 M FeCl₃ at 0.24 T (e) Comparison of respective theoretical velocities from the proposed scaling model with experimental velocities

### 3.5. Scaling analysis of field-induced magneto-thermal advection

Having established that internal advection enhances evaporation rate via shear-induced replenishment of the vapor diffusion layer and that the strength of circulation velocity enhances strongly as a function of magnetic field, the mechanism requires being understood. As discussed [33], the bulb shape of the pendent gives rise to non-uniform evaporation, leading to the generation of thermal as well as the solutal gradient at the droplet surface. This gives rise to surface tension gradient along the surface, leading to interfacial circulation. Further, the adsorption–desorption kinetics of the ions with respect to the bulk and interface of the droplet in conjunction with non-uniform evaporation from the pendent gives rise to thermal and ionic concentration gradient within the bulk of the droplet, leading to internal thermo-solutal advection. In the presence of the magnetic field, the flow visualization exercise shows that the effective internal advection velocity enhances with a typical change in the orientation of the circulation plane. Thereby, it is plausible to expect that such enhanced advection would replenish the vapor layer by interfacial shear to a larger extent, thereby enhancing evaporation. However, a mathematical analysis of the same would cement the mechanistic approach. Further, it also needs to be revealed whether the magnetic field enhances evaporation by modulating the thermal convection or the solutal convection or both. Accordingly, an analysis has been proposed to deduce the dominant mechanism involved in the field influenced augmented evaporation dynamics. The analysis is constrained by the assumptions, viz., surface tension change is solely a function of temperature gradient or concentration difference within the droplet, the droplet only cools down up to the dry bulb temperature of the ambient air while evaporation and thitherto-physical properties do not vary with minor temperature and humidity fluctuations and due to magnetic field.

For the evaporating droplet under the influence of magnetic field, the energy balance at any instant can be expressed as

$$\dot{m} h_{fg} = k_{th} A \frac{\Delta T_m}{R} + \rho C_p U_{c,m} A \Delta T_m + \rho C_p V_f A \Delta T_m \qquad (6)$$

where $\dot{m}$, $h_{fg}$, $k_{th}$, $C_p$ and $\rho$ refer to the change in mass lost as vapour, the enthalpy of vaporization, the thermal conductivity, specific heat and density of the liquid, respectively. $\Delta T_m$, $U_{c,m}$ and $V_f$ represent the temperature difference across the droplet bulk and interface created due to evaporation, the spatially averaged internal velocity (in the absence of magnetic field) and the spatially averaged internal circulation velocity under the influence of magnetic field. The left-hand side of the equation represents the energy flux due to the evaporative loss and right-hand side includes heat conduction across the fluid, convective component of energy which comprises of the thermo-convective circulation and the magneto-thermoconvective circulation, respectively. As the droplet surface cools down due to evaporation, finite difference in temperature is generated across the droplet bulk and the interface, which leads to possible thermal convection within the droplet. Furthermore, uneven cooling of the pendent shape leads to surface convection or thermo-Marangoni convection along the droplet surface [33, 34]. The present approach theorizes that the magnetic field induces enhanced evaporation, thereby leading to augmented thermo-convection, which aids in further hastened evaporation. The average internal circulation velocity due to the thermal gradient generated is expressible as $U_{c,m} = \frac{\sigma_T \Delta T_M}{\mu}$ [34], where $\sigma_T$ represents the partial derivative of the surface temperature as with respect to temperature. Since the flow visualization reveals enhanced circulation, it is assumed that the velocity magnitudes are additive and accordingly Eqn. 6 has been formulated. The expression for the magneto-thermo convection velocity $V_f$ can be deduced by scaling analysis of the magnetohydrodynamic forces within the system. As reported [33], presence of ions in the solvent leads to internal circulation within evaporating droplets due to thermos-solutal advection. Thereby, the droplet evaporation in magnetic field essentially poses as a problem of moving solvated charges in a magnetic field and its implications on the mass transfer process. For a moving charged entity within a coupled electromagnetic field, the net force experienced by the charged body is expressible as

$$\vec{F} = q\vec{E} + (\sigma_e . \vec{E} \times \vec{B}) + \sigma_e (\vec{v} \times \vec{B}) \times \vec{B} \tag{7}$$

In the present condition, the absence of electric field leads to restricting of the Eqn. 7 as

$$\vec{F} = \sigma_e (\vec{v} \times \vec{B}) \times \vec{B} \tag{8}$$

Since the field is assumed to be purely unidirectional (only between the cylindrical poles) and the internal flow has been visualized to occur orthogonally to the field direction, the system can be scaled to yield

$$\vec{F} = \sigma_e U_{c,m} B^2 \tag{9}$$

This additional force experienced by the charged already in motion induces augmented velocity of circulation within the droplet. The force thereby can be expressed in terms of the temporal acceleration ($a$) within the fluid element, leading to the expression

$$\sigma_e U_{c,m} B^2 = \rho a \tag{10}$$

The acceleration can be scaled as $a \sim V_f / t$ (where the time period can be further scaled as $t \sim R/V_f$), thereby leading to the approximate scaled expression as

$$\sigma_e U_{c,m} B^2 \simeq \rho \frac{V_f^2}{R} \tag{11}$$

Accordingly, the approximate expression for the internal circulation velocity under magnetic field influence can be expressed as a function of the zero fields internal circulation velocity as

$$V_f \simeq B\sqrt{\frac{\sigma_e U_{c,m} R}{\rho}} \tag{12}$$

Based on the expression of the internal convection velocity in terms of the thermal gradient generated within the droplet, the augmented velocity under field effect can be expressed as

$$V_f \simeq Ha\sqrt{\frac{\sigma_T \Delta T_m}{\rho R}} \tag{13}$$

In the preceding set of eqns., $q$, $\vec{E}$, $\vec{B}$, $\vec{V}$, $\sigma_e$, $\vec{F}$, $m$, $a$, $t$, and $R$ represent the charge of the solvated ions, electric field intensity, magnetic field strength, velocity, the electrical conductivity of the fluid, electromagnetic force per unit volume, mass and acceleration of the ionic fluid due to field, time and the radius of the droplet. Ha represents the non-dimensional Hartmann number $\left(Ha = BR\sqrt{\frac{\sigma_e}{\mu}}\right)$, the ratio of the electromagnetic Lorentz force to the viscous force in a magnetohydrodynamic system. . Substituting the expressions for $U_{c,m}$ and $V_f$ in eqn. 6 yields

$$\rho \dot{R} R h_{fg} = k_{th}\Delta T_m + R\rho C_p(\frac{\sigma_T \Delta T_m}{\mu})\Delta T_m + R\rho C_p Ha\sqrt{\frac{\sigma_T \Delta T_m}{\rho R}}\Delta T_m \tag{14}$$

Introducing the Marangoni number (Ma) in absence of magnetic field and due to thermal gradients within the droplet as

$$\rho \dot{R} R h_{fg} = k_{th}\Delta T_m(1+ Ma + Ha\sqrt{\frac{R}{\alpha}}\sqrt{\frac{\sigma_T \Delta T_M}{\mu}}\sqrt{\frac{C_p \mu}{K_{th}}}) \tag{15}$$

$$Ma = \frac{R}{\alpha}\sqrt{\frac{\dot{R} h_{fg} \sigma_T}{C_p \mu}} \tag{16}$$

The expression can be further reduced in the form

$$\rho \dot{R} R h_{fg} = k_{th} \Delta T_m (1 + Ma + Ha\sqrt{Ma \Pr}) \tag{17}$$

In eqn. 17, the Ma is >>1 for stable internal circulations to manifest within the droplet [33, 34], and thereby the expression can be further simplified as

$$\frac{\rho \dot{R} R h_{fg}}{k_{th} \Delta T_m} = Ma + \sqrt{HaMa}\sqrt{Ha \Pr} \tag{18}$$

Which can be further condensed as

$$\frac{\rho \dot{R} R h_{fg}}{k_{th} \Delta T_m} = Ma + (Ma_{t,m} \Pr_m)^{1/2} \tag{19}$$

The 2$^{nd}$ term on the right-hand side of eqn. 19 represents the geometric mean of two dimensionless numbers which is absent in the analysis of simple salt based droplet evaporation [33]. In the thermal analysis of magnetic field affected internal circulation, two additional terms, viz. the magnetic Prandtl number ($\Pr_m$=PrHa) as well as the magneto-thermal Marangoni number ($Ma_{T.m}$=MaHa) appear in the mathematical analysis. The nature of these terms essentially signifies that the presence of the magnetic field (represented by the Ha) leads to enhanced internal circulation due to electromagnetic forcing, which enhances the evaporation rate. This, in turn, creates augmented thermal gradients within the droplet, which enhances the thermal advection inside the droplet until a steady state is achieved. Thereby, it is possible to understand the degree to which the thermal advection is modulated due to the field by analyzing the magnitude of the magneto-thermal Ma. It is noteworthy that the generic expression for the Ha is typical for fluids with very high electrical conductivities, such as plasmas. In the present scenario, the magnetic motion in the fluid is majorly caused by the paramagnetic moment of the solvated ions of magnetic materials. Thereby, in the present approach, the values of Ha are obtained employing a re-modified expression for the Ha in terms of magnetic moments, as ($Ha = BMR\rho/u_{cir}\mu$) where $B$, $M$, $R$ and $u_{cir}$ refer to the magnetic field intensity, magnetic moment of the paramagnetic salt, the radius of droplet and the internal advection velocity, respectively. Fig. 6 (b) illustrates the percentage contribution of the different mechanism involved in energy transfer during the evaporation process for paramagnetic salt solution droplets in magnetic field environment. As observed in the preceding analysis, conduction, thermal convection and magneto-thermal convection are three mechanisms which can contribute to energy transfer during the evaporation process. It is noteworthy that the convection term originates due to internal circulation within the solvent phase, brought about by the thermal gradients induced during evaporation. As reported [33], the thermal gradients appear at the droplet surface, as well as the bulk, leading to both surface and bulk advection. However, since the quantification of surface advection by experiments is challenging, the Marangoni number approach enables to correlate it to the bulk advection (from the very definition of the Ma). It is observed from Figure 6 (b) that the conduction transport is equally minimal for all cases. The thermal convection term is dominant, and in presence of magnetic field, the magneto-thermal component is appreciably large. However, the magneto-thermal contribution is observed to be at best ~ 20% of the total. However, the flow visualization study reveals enhancement in internal circulation velocity by 2-3 times, which cannot be sufficiently explained by the magneto-thermal component. Thereby, although potent to induce enhanced evaporation, the magneto-thermal component cannot completely explain the observation and further probing is necessary.

In a similar manner, the circulation velocity induced by buoyancy forces within the droplet is evaluated [34] as

$$u = \frac{g\beta\Delta T_R R^2}{\nu} \quad (20)$$

and the associated Rayleigh number is expressible as

$$Ra = \frac{R^2}{\alpha}\sqrt{\frac{\dot{R}h_{fg}g\beta}{C_p\nu}} \quad (21)$$

where the temperature difference responsible to induce buoyant convection within the droplet can be expressed as

$$\Delta T_R = \sqrt{\frac{\nu \dot{R} h_{fg}}{g\beta R^2 C_p}} \quad (22)$$

In eqn. (20-22), $g$, $\beta$, $\nu$, and $\Delta T_R$ are acceleration due to gravity, coefficient of thermal expansion of the liquid, kinematic viscosity of the liquid and the temperature difference responsible for driving the evaporation process through internal buoyant advection. The possibility of enhanced evaporation in such multi-component droplets can also be linked to Rayleigh convection within the ambient gas phase, however, analysis [34] has shown that it is much weaker compared to the enhancement due to internal circulation in the droplet.

Given the possibility that both thermal Ma and Ra based convection can improve droplet evaporation, the dominant mode needs to be determined. Nield [32] proposed a stability analysis to comprehend convection mechanism in liquids while evaporation. The analysis involves $Ra/$ $Ra_c$ and $Ma/Ma_c$, the ratios of Rayleigh number to critical Rayleigh number and ratio of Marangoni number to critical Marangoni number as

$$\frac{Ra}{Ra_c} + \frac{Ma}{Ma_c} = 1 \qquad (23)$$

The critical Ma and Ra rely on the Lewis number $(Le = q_0 d/k)$ and therefore the variation in evaporation regime regulates the stability pattern of the internal advection. Based on the analysis of Nield [32] and Davis [39], the Ma and Ra values for different salt solution droplets have been plotted for zero magnetic field case in Figure 8. As discussed in literature [33], the addition of salt induces minor amounts of thermal Ma based instability, and the points shift right and upwards on the Ma-Ra plane compared to water. This signifies that internal advection is increasing in case of salts; however, the advection is not conditionally stable as the points still lie below the stability boundaries proposed by both Nield and Davis.

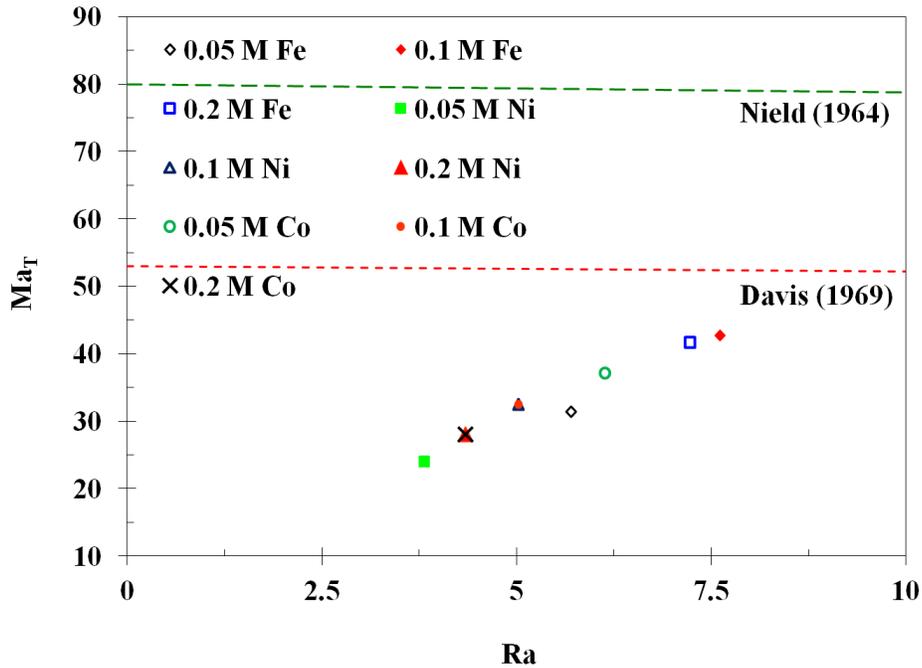

**FIG. 8** Stability plot of the Rayleigh and Marangoni numbers for pure water and different magnetic salt solutions based droplets of varying concentration in the absence of magnetic field. Stability criteria lines proposed by Nield [32] and Davis [39] illustrate the regimes for stable internal advection.

Further, the same stability plot has been presented in case of finite magnetic field environment governing the evaporation process. For the magnetic field case, the effective $Ma_{T,m}$ was evaluated and has been plotted against Ra in Figure 9, similar to Figure 8. Additionally, in the magnetic field case, the stability of points lying in the stability plane will also be governed by the magnetic field strength. Accordingly, iso-Ha lines [19] have been plotted on the $Ma_{T,m}$-Ra stability plane to understand the influence of magnetic field as well. It is observable from the $Ma_{T,m}$-Ra plot in Figure 9 that several salt solution droplets now lie in the regime of critical stability of circulation (in between the boundaries by Davis and Nield). Thereby, the magneto-thermal advection is superior in strength than the zero field thermal advection, and this is experimentally proven by the PIV exercise. The iso-Ha lines reveal that with increase in Ha, the system tends to shift upwards of the stability boundaries; thereby representing that magneto-thermal advection could also be strong enough to induce unconditionally stable circulation within the droplet if the field strength or the magnetic moment is sufficiently high. However, the present points do not go beyond the line of Nield, which represents zone of unconditionally stable circulation. Hence, it cannot be clearly reasoned that magneto-thermal advection is the governing mechanism for enhanced evaporation.

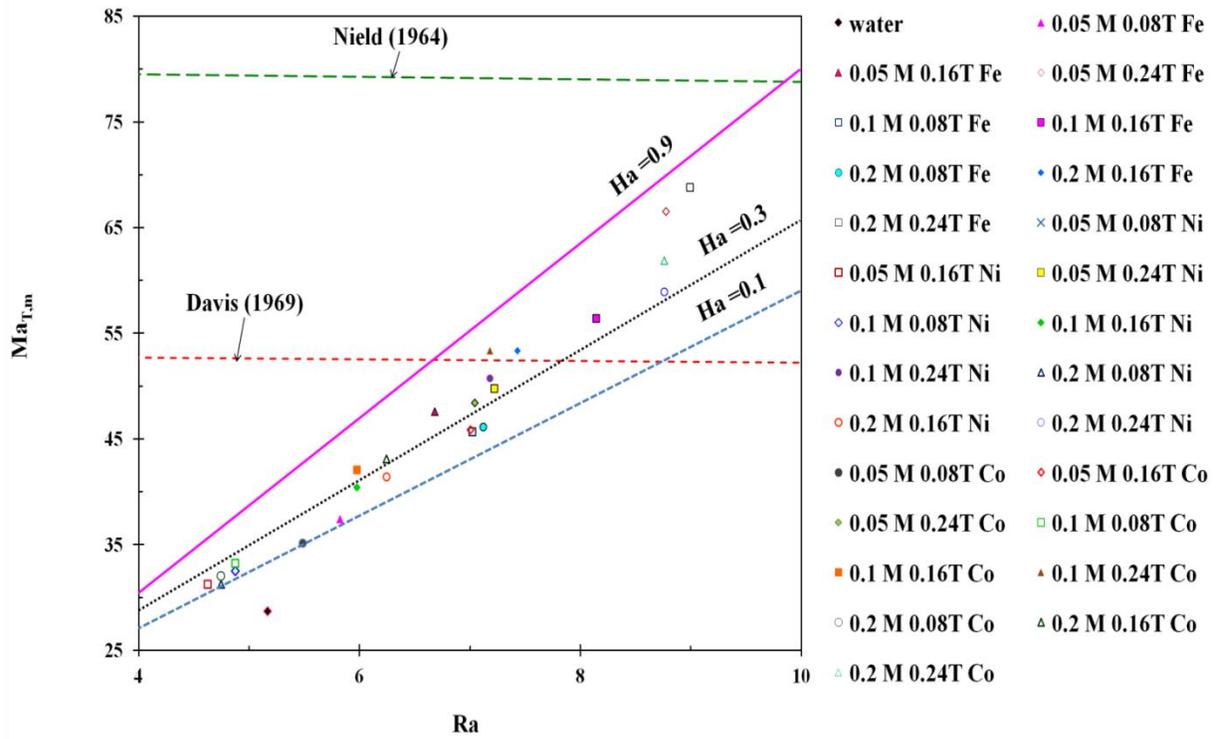

**FIG. 9** Comparison of magneto-thermal Ma and Ra for different magnetic salt solution droplets with varying concentration evaporating under the effect of magnetic field. Iso-Ha lines have been presented to understand the effect of magnetic field strength on the stability regimes.

### 3.6. Governing role of magneto-solutal advection

In multi-component droplets where only one component is evaporating, the effective concentration of the solute phase enhances as the solvent loses mass due to evaporation. In the present case, the paramagnetic ion concentration within the system enhances with the ongoing evaporation of the droplet. Furthermore, solvated ions are surface active agents, which exhibit preferential adsorption or desorption towards the interface or the bulk of the fluid. Thereby, during the evaporation process, a gradient of concentration is expected to exist across the droplet interface and bulk. Furthermore, the pendent shape leads to non-uniform evaporation, leading to the occurrence of concentration gradients at the interface also. Such gradients lead to changes in the value of surface tension (Figure 5 (a)) and thereby are potent to cause interfacial solutal Marangoni advection. Additionally, the gradient across the bulk and interface leads to solutal advection within the droplet. The occurrence of such concentration gradients can be quantified by mapping the dynamic surface tension of the solution. As evaporation progresses, the bulk concentration can be determined from the volume of the pendent (as the product of the volume of fluid and concentration must remain constant). On the contrary, the surface tension of the solution with passing time can be obtained by fitting the instantaneous droplet profile to the Young-Laplace equation. Further, based on the concentration dependent behavior of the surface tension (from Figure 5 (a)), the instantaneous concentration of the ions at the interface can be obtained [33]. Figure 10 illustrates the dynamic concentration evolution at the droplet bulk and the interface for 0.2 M $FeCl_3$ solution at different applied external magnetic fields. It is evident that a concentration difference exists between the bulk of the droplet and that at the interface. This gradient is the driving potential for a solutal convection within the droplet. With the increase in a magnetic field, it can be observed that the gradient enhances, which therefore leads to the magneto-solutal advection, leading to enhanced evaporation.

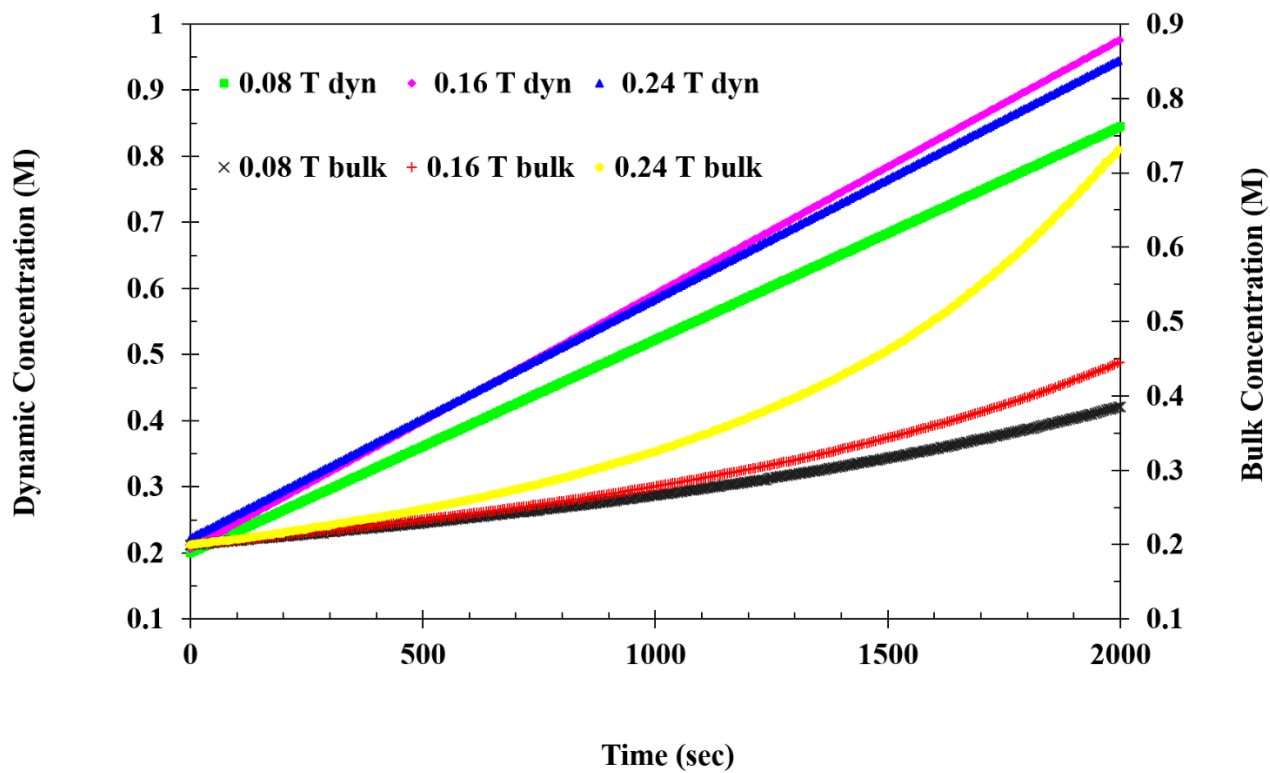

**FIG. 10** The dynamic bulk and interfacial ion concentration of FeCl$_3$ at 0.2 M droplet during the progress of evaporation

The salt induced internal circulation has also been confirmed by Savino and Fico [40]. The reason behind this circulation is primarily caused by the concentration difference of the dispersed phase between the bulk and interface of the droplet. Also as discussed earlier, the pendent shape itself induces solutal Marangoni convection at the interface and this can be correlated to the internal convection via the scaling analysis. A conservation analysis similar to magneto-thermal advection can, therefore, be theorized to understand the role of magneto-solutal advection. Based on species transport from the droplet and within, a species balance equation [33] can be expressed as

$$\dot{m} = DA\frac{\Delta C_m}{R} + U_{c,m} A \Delta C_m + V_{f,c} A \Delta C_m \qquad (24)$$

where $\dot{m}$, $D$, $\Delta C_m$, $U_{c,m}$ and $V_{f,c}$ represent rate of change of mass of the evaporating droplet, coefficient of diffusion of solute in the solvent, difference between the bulk and dynamic interfacial concentrations, internal circulation velocity due to solutal gradient in absence of field and a athe magneto-solutal circulation velocity due to the interplay of the solutal gradient and the applied magnetic field. The internal circulation velocity due to solutal gradient is expressed in a similar fashion as the thermal case, $U_{c,m} = \frac{\sigma_c \Delta C_m}{\mu}$ where $\sigma_c$ represents the rate of change of surface tension with respect to solvated ion concentration. Introducing the expression for the circulation velocity under field effect $V_{f,c}$ (similar as $V_f$ from eqn. 13), the expression yields

$$\rho R \dot{R} = D \Delta C_m + \frac{\sigma_c}{\mu} R (\Delta C_m)^2 + Ha \sqrt{\frac{U_{c,m} \nu}{R}} R \Delta C_m \qquad (25)$$

$$\rho R \dot{R} = D \Delta C_m (1 + Ma_S + Ha\sqrt{Sc}\sqrt{Ma_S}) \qquad (26)$$

Rearranging further and applying the condition that the $Ma_s$ is greater than unity for stable circulation to occur, the expression can be obtained as

$$\frac{\rho R \dot{R}}{D \Delta C_m} = Ma_S + \sqrt{HaSc}\sqrt{HaMa_S} \qquad (27)$$

In eqn. 27, $Sc$ represents the Schmidt number for the solution and $Ma_S$ represents the solutal Marangoni number, expressible as

$$Ma_S = \frac{\sigma_c R \Delta C_m}{D \mu} \tag{28}$$

where $Ma_{S,m}$ and $Sc_m$ portray the magneto-solutal Marangoni number (=HaMa$_s$) and the magneto-Schmidt number (=HaSc) for the system. The inclusion of these two dimensionless number in species balance presents the role played by application of magnetic field in enhanced mass transport during evaporation. Similar to the analysis presented earlier, the percentage contribution of the terms in the magneto-solutal equation has been determined. Figures 11 (a) and (b) illustrate the percentage contribution of all three mechanisms present in the theorized magneto-thermal and magneto-solutal models, respectively. Figure 11 (a) illustrates the contributions of the conduction, zero-field convection and the magneto-thermal convection terms, and for 0.2 M FeCl$_3$ solution at 0.24 T, ~ 20% contribution by the magneto-thermal convective component is observed. On the contrary, in Figure 11 (b), the magneto-solutal convection is observed to contribute ~ 60 % of the net, surpassing the zero-field solutal advection. Thereby, it may at this juncture be argued that the internal convection proposed responsible for the augmented evaporation kinetics is majorly due to the magneto-solutal effect. This reasoning is further cemented by the mathematical deduction of the internal circulation velocity. Based on the experimentally mapped dynamic concentration gradient and the magneto-solutal scaling analysis, the field influenced circulation velocity is obtained from the magneto-solutal Ma. The predicted values have been illustrated against experimental average circulation velocity values in Figure 7 (e) and agreements within 15 % of the experimental values have been achieved. The same analysis when performed employing the magneto-thermal model has been observed to grossly under-predict the velocities. Thereby, accurate velocity deduction from the model validates the proposed magneto-solutal advection theory.

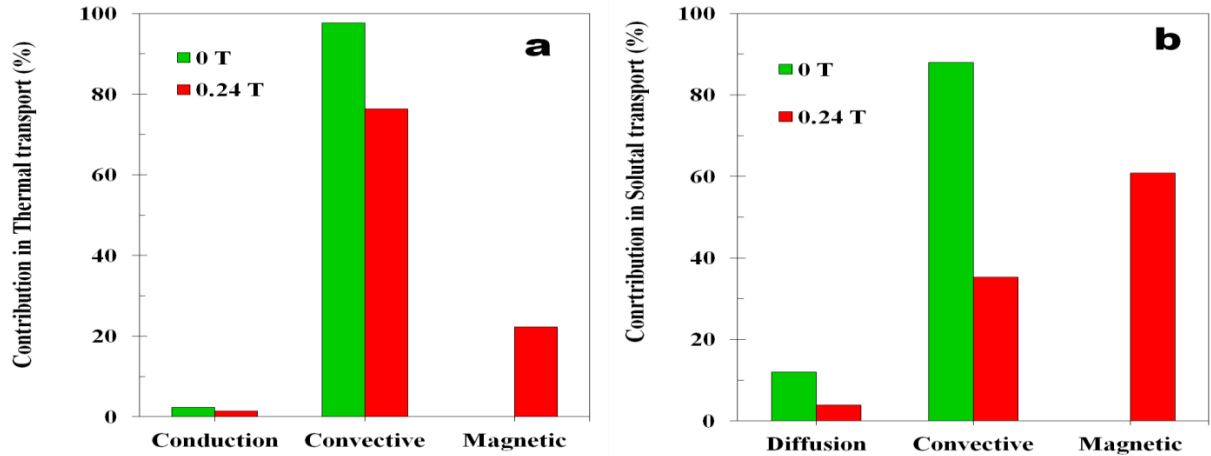

**FIG. 11** (a) The contribution by different mechanisms based on the magneto-thermal convection model (b) The contribution by a different mechanism based on the magneto-solutal convection model.

Having established the magneto-solutal advection as the major mechanism governing the enhanced evaporation dynamics, its degree of dominance over the magneto-thermal advection needs to be understood. Figure 12 illustrates the comparison of thermal Ma with respect to the solutal Ma both in case of both absence and presence of magnetic field. The stability of the internal advection is represented by the iso-Le lines, as proposed by Joo [41]. The points lie far off the Le lines and are greater in magnitude along the $Ma_S$ axis than the $Ma_T$ axis, which represents stable circulation due to solutal gradient and the dominant mode of circulation compared to the thermal mode. It further guides that the effect of solutal Marangoni convection is also more compared to the caliber of thermal Marangoni convection at the droplet interface. The concentration gradient drives the solutal Marangoni advection as well as bulk advection, which leads to faster evaporation due to augmented ionic gradient imbalance between the bulk and interface and at the interface itself. The improved evaporation enhances the evaporative cooling effect, which further affects the thermal

Marangoni convection and the bulk thermal convection. Three different iso-Ha lines for Ha = 0.1, 0.3 and 0.6 are shown to represent the role of the Lorentz force. In the present case it is observed that the thermal and solutal Ma values do not respond much to the magnetic field change, however, the concentration of the solution greatly shifts the points farther away from the origin, thereby representing enhanced stable circulation. This is caused by the solutal advection and is similar to report in the literature [33]. However, in order to quantify the role of the magnetic field, an additional stability plane for the $Ma_{T, m}$ and the $Ma_{S, m}$ has been presented in Figure 13. It can be observed in figure 13 that in comparison to the effects if figure 12, the points drift much farther away with an increase in the magnetic field strength. It can also be observed that the iso-Ha lines drift farther away from the origin compared to figure 12, thereby conveying that the effect of magnetic field essentially pushes the internal circulation towards a zone of greater stability and strength on the stability plane. Thereby, the comprehensive analysis not only shows that the major and dominant governing mechanism is the proposed magneto-solutal advection but also validates the mathematical analysis by quantification of the velocity magnitude and stability plots.

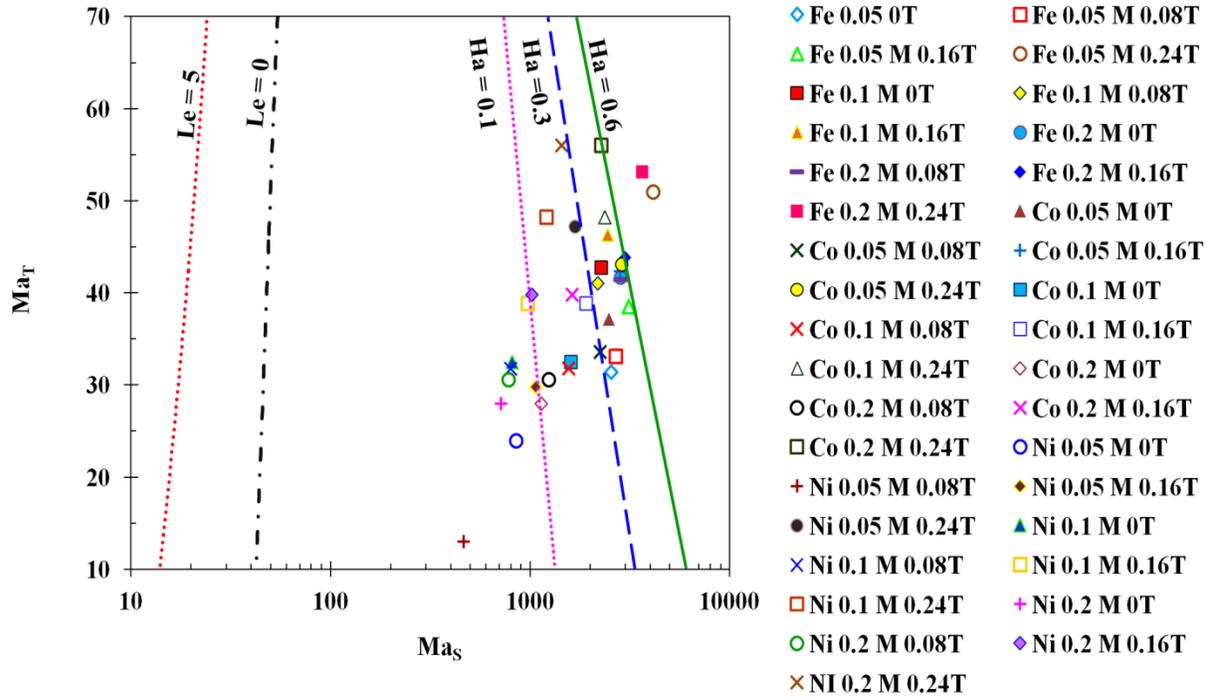

**FIG. 12** Comparison of the thermal Ma against the solutal Ma in absence and presence of magnetic field. The stability curves represent different iso-Le (Joo [41]) and iso-Ha lines.

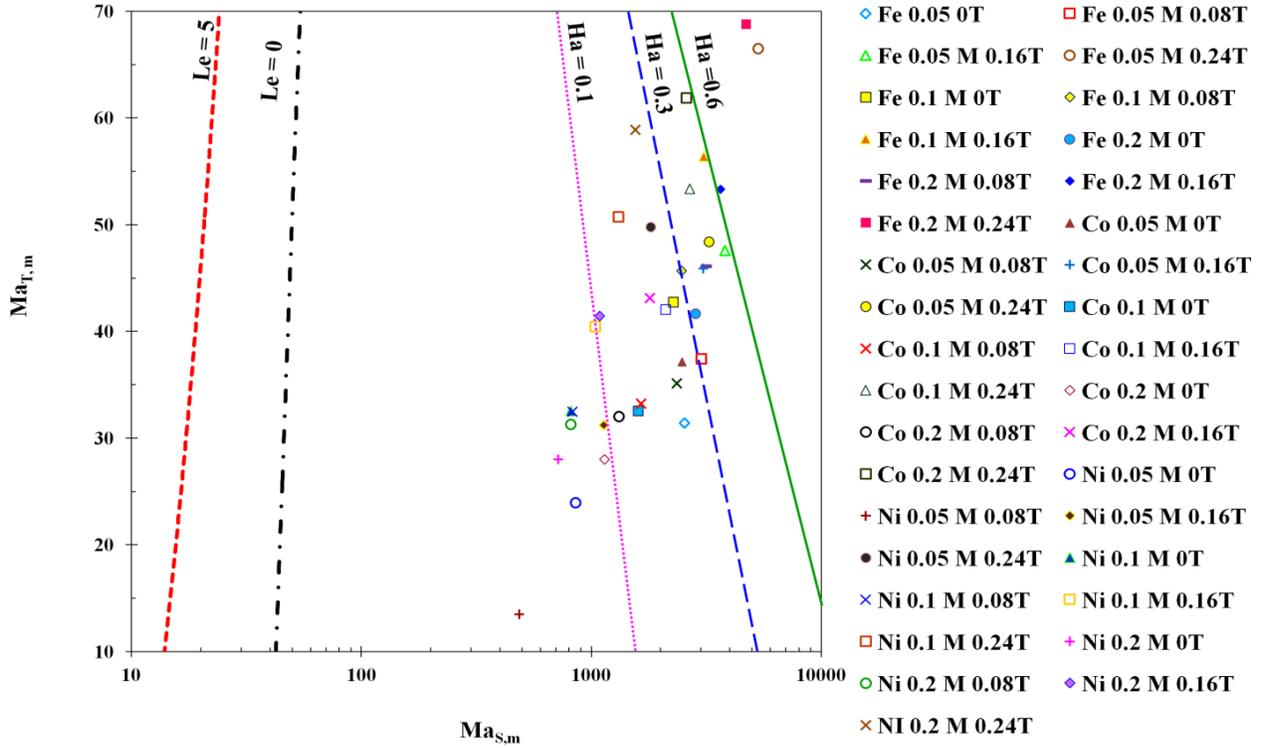

**FIG. 13** Comparison of magneto-thermal Ma against the magneto-solutal Ma. The stability curves iso-Le (Joo [40]) and iso-Ha lines.

## 4. Conclusions

The present article aims at understanding the influence of magnetic field on the evaporation kinetics of paramagnetic solution pendent droplets. Experiments were conducted to study evaporation kinetics of pendant droplet for different magnetic salt solutions under magnetic fields of variant strengths. The classical $D^2$ law is valid but evaporation rate enhancement under influence of magnetic field is observed through experimentation. Literature results point out that presence of solvated ions within the droplet leads to augmented evaporation due to solutal advection within the droplet. The advection current shears the droplet interface, which in turn replenishes the vapor diffusion layer surrounding the droplet with ambient air, thereby improving evaporation. It is further observed from experiments that the degree of enhancement of evaporation is directly proportional to the magnetic moment of the solvated paramagnetic ions, which hints at the role of magnetohydrodynamics mediated internal advection. Particle Image Velocimetry studies reveal that the internal advection velocity enhances with the application of magnetic field and also changes the orientation of rotation. The latter is explained based on the motion of charge in a magnetic field along the lines of Fleming's right-hand rule. Mathematical analysis shows that the diffusion-based evaporation model and surface tension based models are incapable to predict the enhanced evaporation. Accordingly, a scaling based analytical model has been proposed which takes into account the magneto-thermal and magneto-solutal advection within the droplet. The model gives an insight on the insufficiency of the thermal Ma and Ra to produce any appreciable enhancement in internal advection. It is also revealed that the magneto-thermal advection has potential to enhance internal advection; however, it is weak when compared with the magneto-solutal advection. The magneto-solutal advection proves to be the dominant mechanism and the internal velocity magnitudes can be well predicted via the same. Probing into the stability of the magneto-thermal and magneto-solutal Ma at different Ha also reveals the dominant nature of the latter. The enhanced internal convection improves shear on the droplet interface and thereby replenishes the surrounding vapor diffusion layer, leading to augmented evaporation. The present study sheds rich insight into the role of Lorentz forces on solvated paramagnetic ions and their role in modulating the thermo-solutal-magnetohydrodynamics within such microscale droplets and its implications on fluid dynamics, heat, and species transport mechanisms, such as evaporation.

## Acknowledgments

PD thanks IIT Ropar for the financial support towards the present research (vide ISIRD grant IITRPR/Research/193 and IITRPR/Interdisp/CDT).